\documentclass[preprint,12pt]{elsarticle}

\usepackage{lscape}
\usepackage{hyperref}
\usepackage{rotating}
\usepackage{graphicx}
\usepackage{caption}
\usepackage{subcaption}
\usepackage{xr}

\usepackage[utf8]{inputenc} 

\begin{document}
\begin{frontmatter}

\title{Ultra-large alignments using Phylogeny-aware Profiles\tnoteref{label1}}

\author{Nam-phuong D Nguyen\fnref{igb}}
\ead{namphuon@illinois.edu}
\author{Siavash Mirarab\fnref{texas}}
\ead{smirarab@cs.utexas.edu}
\author{Keerthana Kumar\fnref{texas}}
\ead{kk8@cs.utexas.edu}
\author{Tandy Warnow\fnref{igb,uibio,uics}\corref{cor1}}
\ead{warnow@illinois.edu}
\address[igb]{Carl R. Woese Institute for Genomic Biology, University of Illinois at Urbana-Champaign\\
  Urbana, Illinois 61801, USA\\
}
\address[texas]{Department of Computer Science, University of Texas at Austin\\
  Austin, Texas 78712, USA\\
}
\address[uibio]{Department of Bioengineering, University of Illinois at Urbana-Champaign\\
  Urbana, Illinois 61801, USA\\
}
\address[uics]{Department of Computer Science, University of Illinois at Urbana-Champaign\\
  Urbana, Illinois 61801, USA\\
}
\cortext[cor1]{Corresponding author}
\tnotetext[label1]{This paper was selected for oral presentation at RECOMB 2015 and an abstract is published in the conference proceedings.}

\begin{abstract}
Many biological questions, including the estimation
of deep evolutionary histories
and the detection of remote homology between
protein sequences, rely upon multiple
sequence alignments (MSAs) and
phylogenetic trees of
large datasets.
However, accurate large-scale multiple sequence
alignment is very difficult, especially when
the dataset contains
fragmentary sequences.
We present UPP, an MSA method that uses
a new machine learning technique -
the Ensemble of Hidden Markov Models - that we propose here.
UPP produces highly accurate alignments for
both nucleotide and amino acid sequences, even on
ultra-large datasets  or datasets
containing fragmentary sequences.
UPP is available at 
\url{https://github.com/smirarab/sepp}. 
\end{abstract}

\begin{keyword}
multiple sequence alignment \sep phylogeny estimation \sep Hidden Markov models
\end{keyword}
\end{frontmatter}

\section*{Background}

Multiple sequence alignments of large datasets, containing
several thousand to many tens of thousands of sequences,
are used for
gene family tree estimation for multi-copy genes
(e.g., the p450 or 16S genes), the
estimation of viral evolution,
remote homology detection,
the prediction  of the contact 
map between proteins \cite{hwa-pnas},
and the inference of deep evolution \cite{zwickl_increased_2002}; however, 
most
current MSA methods have poor accuracy on large
datasets, especially when they evolved under
high rates of evolution \cite{liu-ploscurrents,dactal}.

\begin{figure}[htpb]
\centering
\begin{center}
\includegraphics[width=1\linewidth]{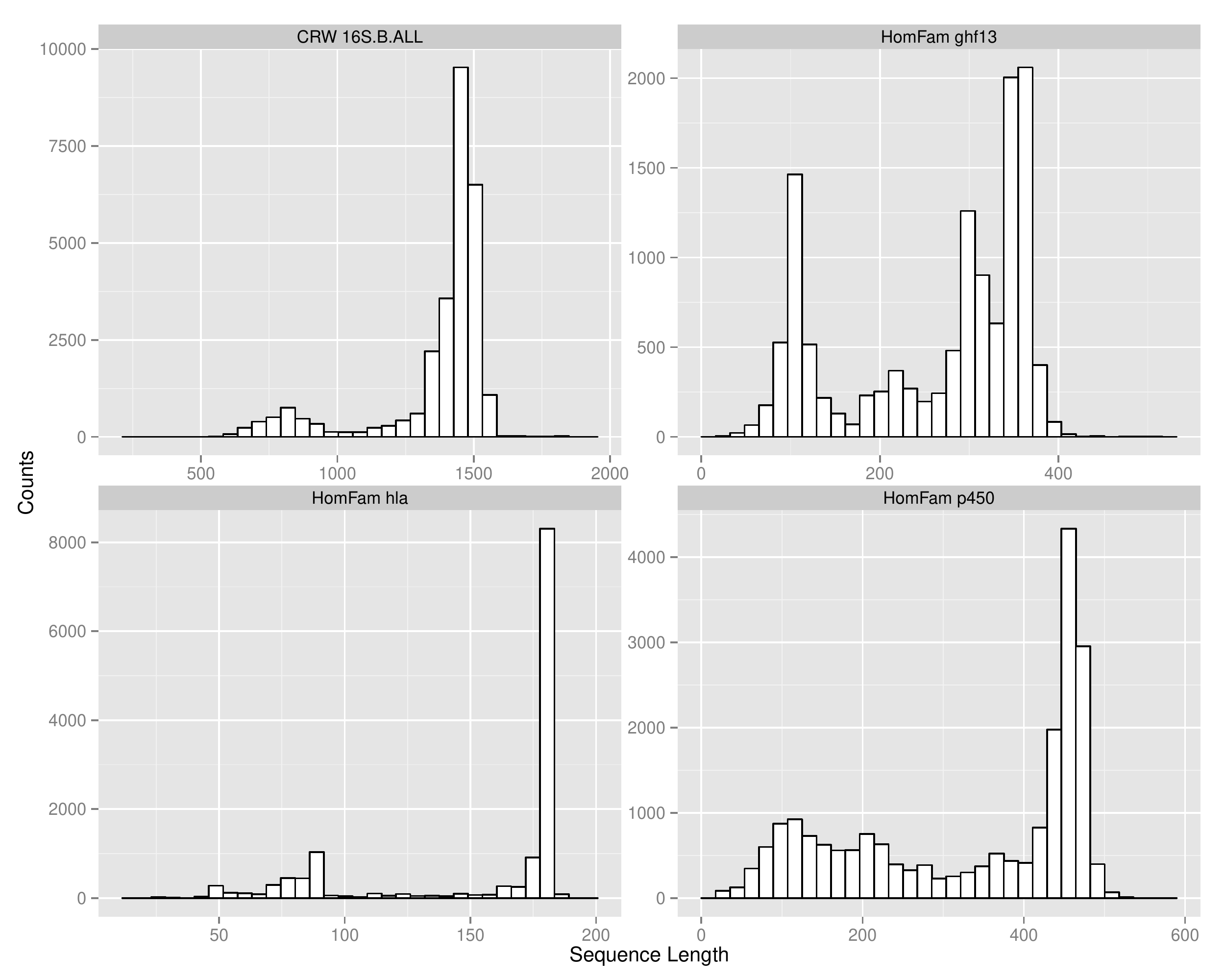}\\
\end{center}
\caption[Histogram of sequence lengths for four fragmentary biological datasets.]{\label{length}  {\bf Histogram of sequence lengths for four of the biological datasets
included in this study.}  These datasets show substantial sequence length heterogeneity and contain a mix of full-length and fragmentary sequences.  
}
\end{figure}

The difficulty involved in
estimating accurate large multiple sequence alignments  
is a major limiting factor for
phylogenetic 
analyses of
datasets containing several hundred sequences or more.
Phylogeny estimation methods that do not 
require a multiple sequence alignment
(e.g., truly alignment-free methods 
\cite{ChanRagan,Ragan2014,DaskalakisRoch} or
almost alignment-free methods such
as DACTAL \cite{dactal}) can be used, 
but alignments are necessary for
the estimation of branch lengths, dates
at internal nodes, the detection of
selection, etc. 
Therefore, 
phylogeny estimation 
generally operates by using methods such as
maximum likelihood (ML) on estimated
multiple sequence alignments.
ML phylogeny estimation
on datasets containing thousands \cite{Stamatakis2014} to tens of thousands
\cite{Price2010}  of
sequences is now feasible, but the accuracy of ML trees depends
on having accurate multiple sequence alignments \cite{Morrison2006}, and
estimating highly accurate large-scale alignments
is extremely challenging; indeed,
some datasets with only 1,000
sequences can be difficult to align with high accuracy \cite{Liu2009,Liu2012}.
Another challenge confronting multiple sequence
alignment methods is the 
presence of fragmentary sequences in the input dataset (see Fig.~\ref{length} for
examples of sequence length heterogeneity found in the biological datasets used in
this study),
which can 
result from a variety of causes, including the use
of next generation sequencing technologies that can produce
short reads that cannot be successfully assembled into
full-length sequences.

We present 
a statistical MSA method that uses
a new machine learning technique that we will
introduce -- the Ensemble of Hidden Markov Models (HMMs) --
to address these limitations.  
Each ensemble of HMMs is best seen
as a collection of profile HMMs for representing a multiple
sequence alignment, constructed  in a phylogeny-aware manner; hence,
we refer to this method as {\em UPP}, for 
{\em Ultra-large alignments using Phylogeny-aware Profiles.}



\begin{figure}[htbp]
\begin{center}
\includegraphics[width=.90\linewidth]{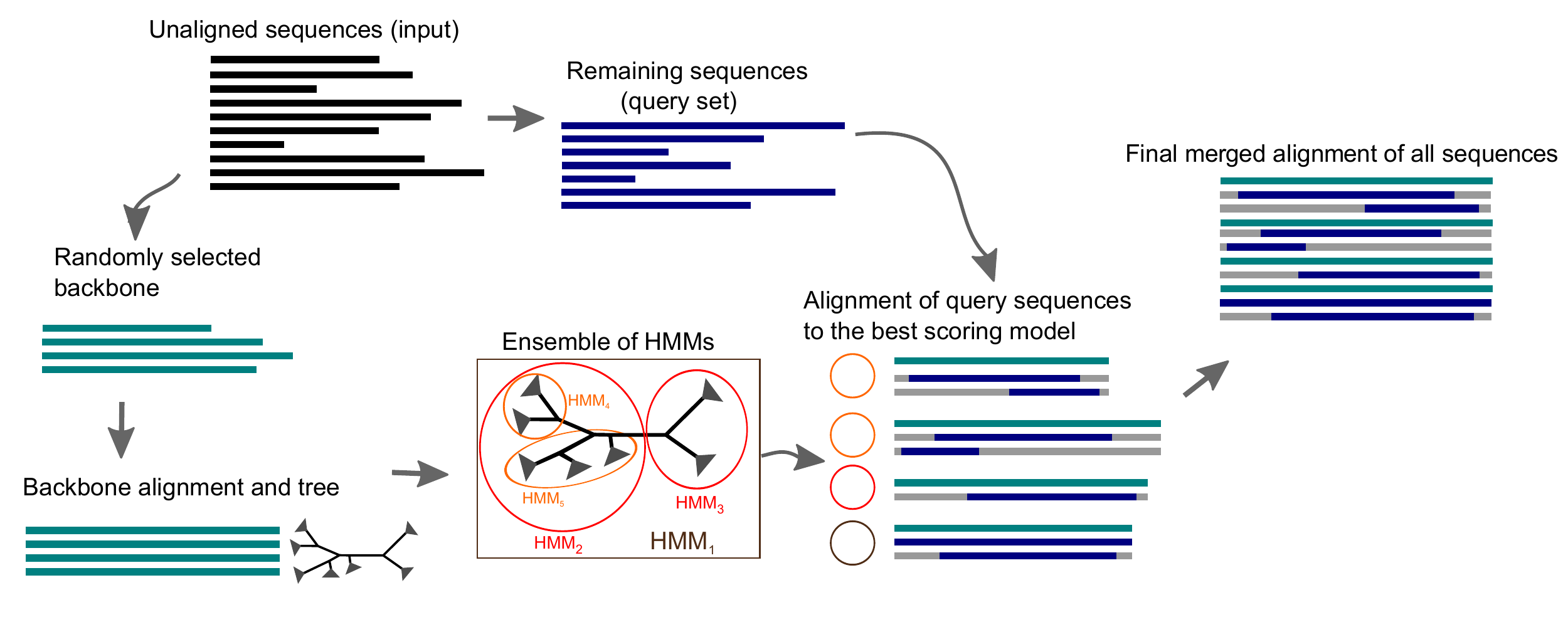}  
\end{center}
\caption{\textbf{Overview of the UPP algorithm.}{\label{flow_chart}}
The input is a set of aligned sequences.
This sequence dataset is 
split into two parts, one the backbone dataset
and the other the  set of query sequences. 
An alignment and tree are estimated
for the
backbone dataset, and an ensemble of HMMs is constructed
based on the backbone alignment and tree. The query sequences 
are then aligned to each HMM, and the best
scoring HMM for each sequence is used to add the
query sequence 
to the backbone alignment.
See text for more details.  }
\end{figure}

UPP uses
the HMMER \cite{HMMER} suite of tools 
(see Methods)  to produce an alignment, and
builds on ideas in SEPP \cite{Mirarab2012}.  The basic idea behind   
UPP is to estimate an accurate alignment on a subset of the sequences and
align the remaining sequences to the alignment using profile Hidden Markov 
Models \cite{Eddy1998}.  
UPP has four phases
(see Fig.~\ref{flow_chart}).  
Phase 1 begins with unaligned sequences and selects a subset 
(called the ``backbone dataset") of the sequences;  
the remaining sequences are the ``query sequences".  
Phase 2 uses PASTA \cite{pasta,pasta-jcb} to compute a multiple
sequence alignment and maximum likelihood tree (which is
unrooted) on 
the backbone sequences; these are called the 
``backbone alignment" and ``backbone tree", respectively.  
As PASTA is a global alignment method and is not designed for the alignment of fragmentary sequences, UPP preferentially selects the backbone sequences from those that are considered to be full-length.  
In order to determine which sequences are ``full-length", UPP only includes 
backbone sequences within 25\% of the length of the typical sequence for the
given locus.  In the case where the typical length of the locus is not known, 
we use the median length of the input sequences as an estimate of the average length  for the 
locus.

This part of the UPP's algorithmic design is similar
to alignment methods that
are based on seed alignments (e.g., 
the technique used in Infernal \cite{infernal}), 
but there is a basic difference between using seed alignments
and these backbone alignments estimated by PASTA.
Seed alignments
are pre-computed  alignments that are typically
highly curated, and may be
based on experimentally verified structural features of the molecule.
UPP does not need to have such seed alignments, and instead is an 
entirely {\em de novo} alignment method.  

Phase 3 creates a collection of 
HMMs (called the ``Ensemble of HMMs'') using the backbone alignment and backbone tree.  The process begins by including the HMM computed on the entire backbone alignment.  
Next, the backbone tree is decomposed by removing a
centroid edge (i.e., an edge that splits the tree 
into two subtrees of approximately equal size).   
For each of these two unrooted subtrees, we use 
hmmbuild (a command within HMMER) to compute an HMM 
on the backbone alignment restricted to the sequences in the leaf set of the subtree,  and then add the resulting HMM to the ensemble.  We repeat this decomposition process until 
each subtree contains at most 10 sequences.  Thus, this process 
results in an ensemble of HMMs, 
each computed on an alignment induced by the backbone
alignment on one of the subtrees.  Note also that while the subtrees are 
local regions within the backbone tree, they may not be clades within
the tree (for example,
in Fig.~\ref{flow_chart}, HMM$_{5}$ is not based upon a clade).  
By default hmmbuild combines nearby sites with more than 50\% gaps into a single match state, making it impossible to form a one-to-one mapping between the match states and the gappy sites in the original subset alignment.  We modify the hmmbuild options to create a match state for each site that has at least one non-gap character, thus making it trivial to map the match states back to the original sites in the subset alignment.  

Phase 4 inserts the remaining query sequences into the backbone alignment, as follows.  The fit of each query sequence to each HMM is assessed using 
hmmsearch (a command within HMMER); this 
returns a bit score, which is a measure of the quality of the match between the query sequence and the HMM. The subset HMM with the best bit score is selected, and the 
sequence is inserted into the subset alignment using hmmalign (a command
within HMMER).  
We treat each site within an
alignment as a statement of positional homology, so that all
letters within the site are considered to be positionally homologous
\cite{fastsp}.
Since positional homology is an equivalence
relation (i.e., a binary relation  that is reflexive, symmetric,
and  transitive),
by transitivity,
this process defines how the query sequence should 
be added into the backbone alignment; similar uses of transitivity have
been used in other multiple sequence alignment methods
\cite{PROMALS3D,pasta-jcb}.
When the sequence has a letter  (nucleotide or amino acid) that is not aligned to any letter in the backbone alignment, the extended alignment will have an ``insertion site".

Once all the query sequences are added into the backbone alignment, 
transitivity defines the final output multiple sequence alignment.   
This approach will tend to have potentially many insertion sites;
in order to save space, we combine adjacent insertion sites into a single column.  
These introduced columns therefore contain nucleotides or amino acids that
are not homologous to each other, and so the columns are
indicated as insertion sites and masked before running a phylogenetic analysis. 
We also do not consider the homologies within these columns when 
evaluating  the accuracy of computed  alignments. 

As we will show, UPP
provides very good accuracy on both phylogenetic and
structural benchmarks, and  is fast enough to 
produce highly accurate alignments on 10,000 sequences in under an hour,
and on one million sequences in twelve days, using
only 12 cores.
Furthermore, UPP has excellent accuracy 
even when the sequence dataset contains a large number of
highly  fragmentary sequences.
In comparison, most other multiple sequence alignment methods
either cannot analyze datasets of the same size due to
computational limitations, or do not exhibit the same
accuracy as UPP under the 
most challenging conditions (large datasets with fragmentary
sequences). 

\section*{Results and discussion}
We used 
a variety of simulated and biological datasets 
from prior publications to compare UPP to
existing multiple sequence alignment methods (see Methods for
details). The
simulated datasets include
{\bf ROSE NT:}  a
collection of 1000-sequence nucleotide datasets;
 {\bf Indelible 10K:}  a collection of
10,000-sequence
nucleotide datasets;
{\bf
RNASim:}  a collection of
datasets ranging from
 10,000 sequences to 1,000,000 sequences;
and
{\bf ROSE AA:} a collection of
5000-sequence simulated AA datasets.
The biological datasets include
{\bf  CRW:}  the three largest
datasets (16S.3, 16S.T, and 16S.B.ALL) from the
Comparative Ribosomal Website   \cite{Cannone2002}
with up to 27,643 sequences; 
{\bf 10 AA:}  ten
amino acid
datasets  with curated multiple sequence alignments
with up to 807 sequences;  
and 
{\bf HomFam:}  
19 large HomFam datasets \cite{Sievers2011}, with up to 
93,681 sequences.
For some of these datasets, 
we generated fragmented versions,
making 12.5\%, 25\%, and 50\% of the
sequences fragmentary, in order to 
evaluate robustness to fragmentary data. 
The simulated datasets have true alignments and trees available
from the prior publications. 
The biological datasets have reference alignments based on
structural features, and the CRW and 10AA datasets also have
reference trees computed using the reference alignments, which
are also available from the prior publications.
The reference alignments for the HomFam datasets 
are too small (5-20 sequences, median 7) and 
trees computed on these reference alignments were
too poorly supported to be useful 
for evaluation purposes.

We computed ML trees
on the estimated alignments, and
report  tree error using the false
negative (FN) rate (also known as the missing branch rate),
and the $\Delta$FN rate, which is
the difference between the FN rates of trees computed on
estimated and true or reference alignments. 
We report 
alignment SP-error, which is the average of the sum-of-pairs
false negative (SPFN) and false positive (SPFP)
rates \cite{fastsp}. We also report the total column score (TC), which
is percentage of aligned columns (i.e., columns with at least
one homology) in the true or reference alignment
that appear in the estimated multiple sequence 
alignment.  

\paragraph{UPP algorithm design. }

We explored modifications of the UPP design in which
we varied the backbone size, used a single HMM instead of an ensemble, 
built ensembles based on clades within the backbone tree,  
built ensembles based on disjoint subsets of ten sequences each, 
used different MSA methods to compute the backbone alignment,
used MAFFT instead of hmmalign to add sequences to the backbone alignment, 
and ran hmmbuild using different
options to compute HMMs on each subset alignment. 
These preliminary studies revealed the following trends.

(1) Using small backbones (100 sequences) rather than large backbones
(1000 sequences) typically produced higher alignment SP-error rates and
tree error rates for both the Ensemble 
of HMMs approach and the single HMM approach (SOM Section \ref{upp_sepp_hmm}).
Using smaller backbones reduced the
running time  for the Ensemble of HMMs approach and had negligible impact on
the running time for the single HMM approach
(SOM Section \ref{upp_sepp_hmm}).

(2) Using an ensemble of HMMs
rather than a single HMM with 1000-sequence backbones
had varying impact. As shown in
Table \ref{table1-v2}, the impact on
alignment SP-error 
ranged from neutral (changes of at most 0.3\% for alignment SP-score
or tree error) to beneficial; for example,
using an ensemble of HMMs had 23.0\% alignment SP-error
on the HomFam datasets whereas using a single HMM produced
alignment SP-error of 25.4\% (Table \ref{table1-v2}).
The impact on TC score also varied: TC scores were better using
single HMMs for the Indelible simulated datasets, and were otherwise
better using
ensembles (Table \ref{table1-v2}). 
The differences in TC score were generally small (e.g., the average difference
was less than 0.5\%).
On the HomFam datasets,  using an ensemble
of HMMs had TC score of 46.6\% while
a single HMM had TC score of 44.5\% (a difference of
2.1\%) and on the Indelible 10000M4 datasets using
a single HMM had TC score of 30.5\% and using
an ensemble of HMMs had 27.4\% (a difference of 3.1\%).
Finally, the use of an ensemble of HMMs instead of a single HMM
generally  reduced tree error (Table
\ref{table1-v2}). For example, results on
the CRW datasets show that using an ensemble of HMMs had average
tree error of 7.8\%, but using a single HMM had
average tree error of 16.5\% (i.e., more than double the tree error). Substantial
reductions in tree error were also observed for the RNASim 10K datasets. In
a few cases (i.e.,  the ROSE AA and Indelible datasets), using a single HMM improved tree
error, but the differences were very small  (Table \ref{table1-v2}). 
The impact of using an ensemble of HMMs instead of a single HMM
was lessened for 100-sequence backbones, and in some cases even led to
small improvements
(SOM Section \ref{upp_sepp_hmm}).
However, the best results were still obtained using the
1000-sequence backbones
with the ensemble of HMMs.

(3) Using ensembles of HMMs computed for
clades within the backbone tree
produced alignments and trees that 
were generally as accurate (according
to the SP-error and tree error rates) and had variable impact on
TC scores  (generally reducing scores but
in some cases improving them)  as those produced using ensembles based
on the centroid-edge decompositions (SOM Section \ref{upp_sepp_hmm} and SOM Table~\ref{table1-v2-supp}). However,
UPP using  clade-based ensembles took more time
(SOM Section \ref{upp_sepp_hmm}).

(4) Using ensembles of HMMs based on disjoint subsets (each with at most 10 sequences) 
had variable impact. For many datasets
(e.g., the ROSE AA, RNASim, CRW, and HomFam  datasets) the impact 
of using disjoint subsets was very small, and in some cases even slightly favorable
(SOM Section \ref{upp_sepp_hmm} and SOM Table~\ref{table1-v2-supp}).
However, for some other datasets 
using disjoint subsets greatly reduced 
accuracy.
For example, on the Indelible 10000M2 datasets, 
default UPP had
alignment SP-error of 3.5\%,  TC score 1.2\%, and $\Delta$FN  error of 0.6\%,
but using disjoint subsets had
SP-error of
28.2\%, TC score 0.3\%, and $\Delta$FN tree error of 19.9\%
(SOM Table~\ref{table1-v2-supp}).
Thus, although
using disjoint ensembles of HMMs reduced the
running time (SOM Section \ref{upp_sepp_hmm}),
  the default ensemble of HMMs was a more reliable technique than
ensembles based on disjoint subsets.

(5) The technique used to estimate the backbone alignment
had a large impact on the final alignment and tree 
(SOM Section \ref{backbone_method}), so that
the final alignment 
SP-error very closely matched the initial backbone alignment SP-error (SOM Section~\ref{backbone_alignment_error}).
Hence, the best alignment methods are needed to
produce the backbone alignment.  

(6)  Using MAFFT to add sequences to the
backbone alignment instead of UPP's default technique (hmmalign, a command
within HMMER) reduced accuracy (SOM Section~\ref{query_method}). 

(7) Using different hmmbuild options (such as turning off the 
entropy-weighting flag) did not improve accuracy (SOM Section  \ref{hmmbuild_variant}).

Overall, 
 the most reliable results were obtained
using large backbones (1000 sequences), using an ensemble of
HMMs, computing the backbone using PASTA, and using
hmmalign to add sequences into the backbone alignment.  These
settings were used for the default version of UPP.
However, for running-time purposes
(so that  ultra-large datasets can be analyzed 
quickly), we explore 
UPP(Fast), the variant of UPP that uses backbones of 100 sequences but
otherwise uses all the default settings
(i.e., restrict the backbone to full-length sequences, 
use an ensemble of HMMs, use PASTA to align
subsets, etc.).

\begin{table}[htpb]
\caption[Comparison of two UPP variants on representative full-length datasets.]{\label{table1-v2}  {\bf Comparison of two UPP variants on 
representative full-length datasets
with respect to alignment SP-error, tree error, and TC scores.}  
All criteria (errors and scores) given as percentages. 
See text for explanation of names of methods and computational
platforms used.
The default setting for UPP is denoted
UPP(Default); it uses a backbone of size 1000, uses PASTA
to compute the backbone alignment, and the ensemble of HMMs technique.
The ``NoDecomp" versions of these two methods replace the ensemble of HMMs
technique with a single HMM.  
Maximum likelihood trees are estimated using RAxML (on the 10 AA datasets) or FastTree (all other datasets) except for HomFam, 
where we do not estimate ML trees as there are no reference trees for the HomFam datasets.}
\centering
\scalebox{.75}{
\begin{tabular}{|l|l|r|r|r|}
  \hline
Model condition & Method & Alignment SP-error & $\Delta$FN & TC score \\ 
  \hline
  10 AA & UPP(Default) & 24.2 & 3.4 & 11.4 \\ 
  10 AA & UPP(Default,No Decomp) & 24.5 & 5.2 & 11.0 \\ 
  \hline
  ROSE AA & UPP(Default) & 2.9 & 1.8 & 2.6 \\ 
  ROSE AA & UPP(Default,No Decomp) & 2.8 & 1.4 & 2.5 \\ 
  \hline
  CRW & UPP(Default) & 12.5 & 7.8 & 1.4 \\ 
  CRW & UPP(Default,No Decomp) & 13.3 & 16.5 & 0.9 \\ 
  \hline
  HomFam(19) & UPP(Default) & 23.0 & NA & 46.6 \\ 
  HomFam(19) & UPP(Default,No Decomp) & 25.4 & NA & 44.5 \\ 
  \hline
  Indel. 10000M2 & UPP(Default) & 3.5 & 0.6 & 1.2 \\
  Indel. 10000M2 & UPP(Default,No Decomp) & 3.3 & 0.5 & 1.4 \\
  \hline
  Indel. 10000M3 & UPP(Default) & 1.3 & 0.2 & 4.6 \\
  Indel. 10000M3 & UPP(Default,No Decomp) & 1.3 & 0.1 & 4.8 \\
  \hline
  Indel. 10000M4 & UPP(Default) & 0.3 & $<$0.0 & 27.4 \\
  Indel. 10000M4 & UPP(Default,No Decomp) & 0.5 & $<$0.0 & 30.5 \\
  \hline
  RNASim 10K & UPP(Default) & 9.5 & 0.8 & 0.5 \\ 
  RNASim 10K & UPP(Default,No Decomp) & 11.2 & 3.0 & 0.3 \\ 
  \hline
\end{tabular}}
\end{table}

\paragraph{Comparison to other MSA methods on full-length sequences. }\label{comparison}
We used
Clustal-Omega \cite{Sievers2011},
MAFFT \cite{Katoh2007},
Muscle \cite{Edgar2004a},
PASTA \cite{pasta,pasta-jcb},
and UPP to compute multiple sequence alignments.

We rank methods by tiers, where the first tier contains 
the method that had the best performance as well as any other 
method that was within 1\% of the best result on the dataset.  Similarly,
the second tier contains the method not in the first
tier that had the best 
performance,  and all methods within 1\% of that method (and
so forth for the remaining tiers).
The method that had the best performance
overall within a collection is also identified.
We describe the general performance of each method on 
the full-length datasets (Table \ref{table:all_full}) and fragmentary 
datasets (Table \ref{table:all_frag}).  
For the fragmentary results, we take the average 
performance of each method on the entire range of fragmented datasets.

The majority of experiments were run on the
homogeneous Lonestar  cluster at
the Texas Advanced Computing Center (TACC).
Because of limitations imposed by Lonestar, these
analyses are limited to 24 hours, using
 12 cores with 24 GB of memory;
methods that failed to complete within 24 hours or
terminated with an insufficient memory error message were marked as failures.
For experiments on the million-sequence RNASim dataset,
we ran the methods on a
dedicated machine with 256 GB of main
memory and 12 cores and ran until an alignment
was generated or the method failed.
We also performed a limited number of experiments on TACC with
UPP's internal checkpointing mechanism, to explore performance
when time is not limited.
All methods other than  Muscle had parallel implementations and 
were able to take advantage of the 12 available cores.

On full-length datasets 
 (Table~\ref{table:all_full}) where
nearly all methods were able to complete, 
PASTA was nearly
always in the first tier with respect
to alignment SP-error, tree error, and TC scores
(the only exceptions being the RNASim 10K datasets,
where PASTA came in second tier for alignment SP-error
and the HomFam(17)  datasets where PASTA came
in second tier for TC score). 
UPP(Default) had the second best
performance: 
in the first tier in terms of SP-error
except for  the Indelible 10K and
HomFam(2) datasets,  where  it was in the
second tier
(with 1.2\% and 3.4\% higher error than the best method),
in first or second  tier for tree error,
and in the first through third tiers for TC score.
MAFFT was in third place, placing in the first through
third tiers for alignment SP-error, first through third
tiers for tree error, and first through fourth
tiers for TC scores.
Muscle and Clustal-Omega came in behind
MAFFT. Muscle 
came in second through fourth tiers
with respect to alignment SP-error, first through
fourth tiers with respect to tree error, and
second through fourth tiers with respect to
TC score.
Clustal-Omega came in first through fourth tiers
with respect to alignment SP-error, second through
fourth tiers with respect to tree error, and
first through fourth tiers with respect to TC scores.
In general, the relative performance of
Muscle and Clustal-Omega seemed to depend on the
type of data, with Muscle doing better on the nucleotide
datasets and Clustal-Omega doing better on the
amino acid datasets.

Thus, for full length sequences,
whether with respect to alignment SP-error, tree error, or TC
scores,  on average
PASTA came in first, UPP came in second, MAFFT
came in third, and Muscle and Clustal-Omega came in
behind these methods.

\begin{table}[htbp]
\caption[Average  alignment SP-error, tree error, and TC score, across most full-length datasets.]{\label{table:all_full}
{\bf Average alignment SP-error,  tree error, and TC score across most full-length datasets. }  
We report the average alignment SP-error  (the average of SPFN and SPFP error) (top),
average $\Delta$FN error (middle), and average TC score (bottom), 
on the collection of full-length datasets.  
All scores represent percentages, and
so are out of 100.
Results marked with an ``X'' indicate that the method failed to terminate within the time limit (24 hours on a 12 core machine).  Muscle failed to align two of the HomFam datasets; 
we report separate average results on the 17 HomFam datasets for all methods and the two HomFam datasets for all but Muscle.
We did not test tree error on the HomFam datasets (therefore, the $\Delta$FN error is indicated by ``NA'').  The tier ranking for each method is shown parenthetically.}
\centering
\scalebox{0.70}{
\begin{tabular}{|l|r|r|r|r|r|r|r|r|}
  \hline
Method & ROSE  & RNASim  & Indelible  & ROSE  & CRW & 10 AA &HomFam & HomFam\\ 
& NT & 10K & 10K & AA & & &(17) & (2) \\
  \hline
 \multicolumn{9}{|c|}{Average Alignment SP-Error}\\
  \hline
  UPP  & 7.8 (1)  & 9.5 (1)&  1.7  (2)&  2.9 (1)& 12.5 (1)& 24.2 (1)& 23.3 (1)&20.8 (2)\\   
  PASTA         & 7.8 (1)  & 15.0 (2)& 0.4  (1)& 3.1 (1)& 12.8 (1)& 24.0 (1)&  22.5 (1)&17.3 (1)\\   
  MAFFT         & 20.6 (2) & 25.5 (3)& 41.4 (3)& 4.9 (2)& 28.3 (2)& 23.5 (1)&  25.3 (2)&20.7 (2)\\ 
  Muscle        & 20.6 (2) & 64.7 (5)& 62.4 (4)& 5.5 (3)& 30.7 (3)& 30.2 (2)&  48.1 (4)&X       \\   
Clustal   & 49.2 (3) & 35.3 (4)& X       & 6.5 (4)& 43.3 (4)& 24.3 (1)&  27.7 (3)&29.4 (3)\\ 
   \hline
 \multicolumn{9}{|c|}{Average $\Delta$FN Error}\\
  \hline
  UPP  & 1.3 (1) & 0.8 (1) & 0.3  (1) & 1.8 (1)    & 7.8  (2)& 3.4 (2)& NA& NA\\   
  PASTA         & 1.3 (1) & 0.4 (1) & $<$0.1 (1)& 1.3 (1)   & 5.1  (1)  & 3.3 (1)& NA& NA\\   
  MAFFT         & 5.8 (2) & 3.5 (2) & 24.8 (3)& 4.5 (3)   & 10.1 (3)  & 2.3 (1)& NA& NA\\   
  Muscle        & 8.4 (3) & 7.3 (3) & 32.5 (4)& 3.1 (2)   & 5.5 (1)   & 12.6(3)& NA& NA\\ 
  Clustal & 24.3 (4)& 10.4 (4)& X        & 4.2(3)   & 34.1 (4)  & 3.5 (2)& NA& NA\\ 
   \hline
 \multicolumn{9}{|c|}{Average TC score}\\
  \hline    
  UPP           & 37.8 (1)    & 0.5  (2)     & 11.0   (3)     & 2.6   (2)  & 1.4  (1)    & 11.4  (1)& 47.3 (1)& 40.3  (3)\\   
  PASTA         & 37.8 (1)    & 2.3  (1)     & 48.0   (1)    & 5.4  (1)   &  2.3  (1)    & 12.1  (1)& 46.1 (2)& 50.0 (1)  \\   
  MAFFT         & 31.4 (2)    & 0.4  (2)     & 7.8    (4)     & 0.6  (3)   & 0.7  (2)    & 12.1 (1) & 45.5 (2)& 46.9 (2) \\   
  Muscle        & 9.8  (3)    & $<$0.0 (2)   & 18.3   (2)   & 2.7  (2)   & 0.7  (2)      & 10.5  (2) & 27.7(4)& X     \\ 
  Clustal       & 5.7  (4)    & 0.2   (2)    & X            & 3.1  (2)   & 0.1 (2)       & 11.8  (1)& 38.6 (3)& 31.0  (4)\\  
   \hline   
\end{tabular}}
\end{table}

\paragraph{Comparison to other methods on datasets with fragmentary sequences. }
We next investigated performance on datasets with fragmentary
sequences.  
As shown in Table \ref{table:all_frag}, UPP was in the first
tier of methods on all the fragmentary datasets with
respect to alignment SP-error,
and in the first tier of methods for three of the four 
collections (except for CRW) with respect to tree error, where
it is in the second tier. 
PASTA was not in the first tier for any collection with respect to
either criterion, and was instead in the second through fourth tiers
for alignment SP-error and second and third tiers for tree error.
MAFFT was in the second and third tiers for alignment SP-error,
but did reasonably well for tree error: 
in the first tier for CRW and otherwise second and third tiers. As before,
Muscle and Clustal-Omega did less well than the other methods: in
the third through fifth tiers, and Clustal-Omega was unable to
analyze at least one dataset.
Note also that the absolute error generally increased,
and that only UPP had reasonably low alignment SP-error and tree error across
all these fragmentary datasets. 
Thus, the relative and absolute performance of methods changed between the
full-length and fragmentary data.

\begin{figure}[htpb]
 \centering
 \includegraphics[width=.95\linewidth]{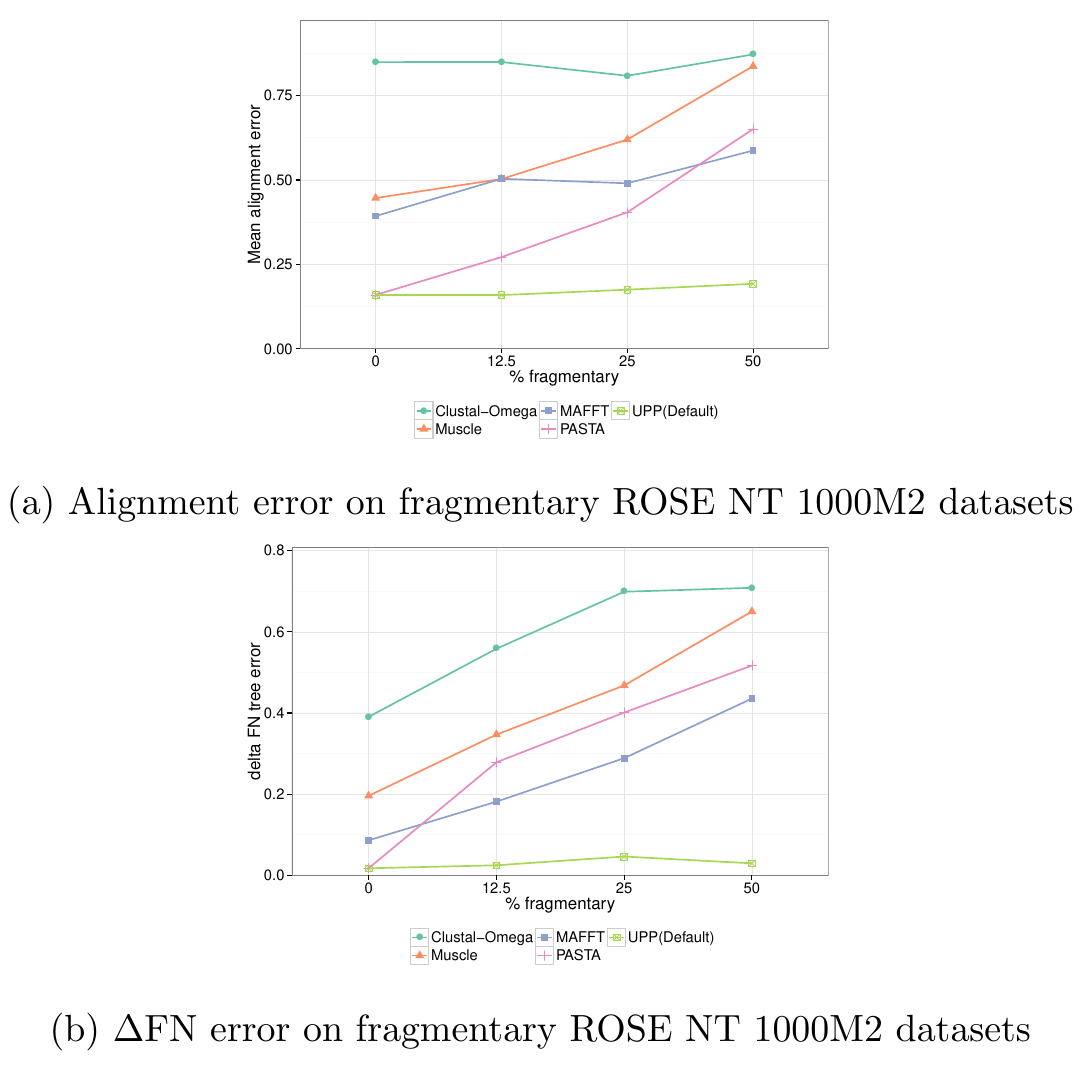}  
 \caption[Impact of fragmentary sequences on alignment SP-error and tree error on the ROSE NT 1000M2 dataset.]{\label{fig:fragmentary_1000M2}  
 {\bf Impact of fragmentary sequences on alignment SP-error and tree error. }
 We show alignment and tree error rates for different methods on the ROSE NT 1000M2 datasets, but include results where a
 percentage of the sequences are made fragmentary, varying
 the percentage from 0\% to 50\%.
 Fragmentary sequences have average 
 length 500 
 (i.e., roughly half the average sequence length for ROSE 1000M2).} 
 \end{figure}

To examine the impact of fragmentation in detail, see
Figure
\ref{fig:fragmentary_1000M2}, which
shows results on the ROSE NT 1000M2 (a very
challenging condition due to high rates of indels and
substitutions), with varying levels of fragmentation.
UPP's alignment SP-error increased only slightly with
increases in fragmentation,  
even  up to the highest degree of fragmentation (50\%).
All other methods exhibited greater increases in
alignment SP-error or tree error than UPP, as the
amount of fragmentation increased.

\begin{table} [htbp]
\caption[Average alignment SP-error and tree error across fragmentary datasets.]{\label{table:all_frag}
{\bf Average alignment SP-error and tree error across fragmentary datasets. }  
We report the average alignment error (top) and average $\Delta$FN error (bottom) 
on the collection of fragmentary datasets.  Clustal-Omega failed to align any of the Indelible 10000M2 fragmentary datasets and thus we mark the results with an ``X''.  
The tier ranking for each method is shown in parentheses.
}
\centering
\scalebox{.80}{
\begin{tabular}{|r|r|r|r|r|}
  \hline
Method & ROSE NT & RNASim 10K & Indelible 10K &CRW \\ 
& & & & (16S.3 and 16S.T) \\
\hline 
\multicolumn{5}{|c|}{Average Alignment SP-Error}\\
  \hline
  UPP & 8.3 (1)  & 11.8 (1) & 2.7 (1)   &16.1 (1)\\   
  PASTA        & 25.2 (2) & 47.7 (4) & 8.8 (2)   &23.3 (2)\\   
  MAFFT        & 32.5 (3) & 25.5 (2) & 51.3 (3)  &24.5 (3)\\   
  Muscle       & 35.3 (4) & 82.2 (5) & 77.6 (4)  &70.6 (5)\\   
 Clustal & 62.0 (5) & 35.0 (3) &  X        &46.7 (4)\\ 
   \hline
 \multicolumn{5}{|c|}{Average $\Delta$FN Error}\\
  \hline
  UPP    & 1.9  (1) & 3.1 (1) &  2.5 (1)   &7.4 (2)\\   
  PASTA           & 25.2 (3) & 21.9 (3)& 9.0 (2)    &8.2 (2)\\   
  MAFFT           & 18.0 (2) & 6.2 (2) &  35.6 (3)  &2.5 (1) \\   
  Muscle          & 27.5 (4) & 43.6 (5)& 45.2 (4)   &30.1 (3)\\   
  Clustal   & 47.8 (5) & 26.3 (4)& X          &37.4 (4)\\ 
   \hline
\end{tabular}}
\end{table}

\begin{figure}[htpb]
\centering
\includegraphics[width=.65\linewidth]{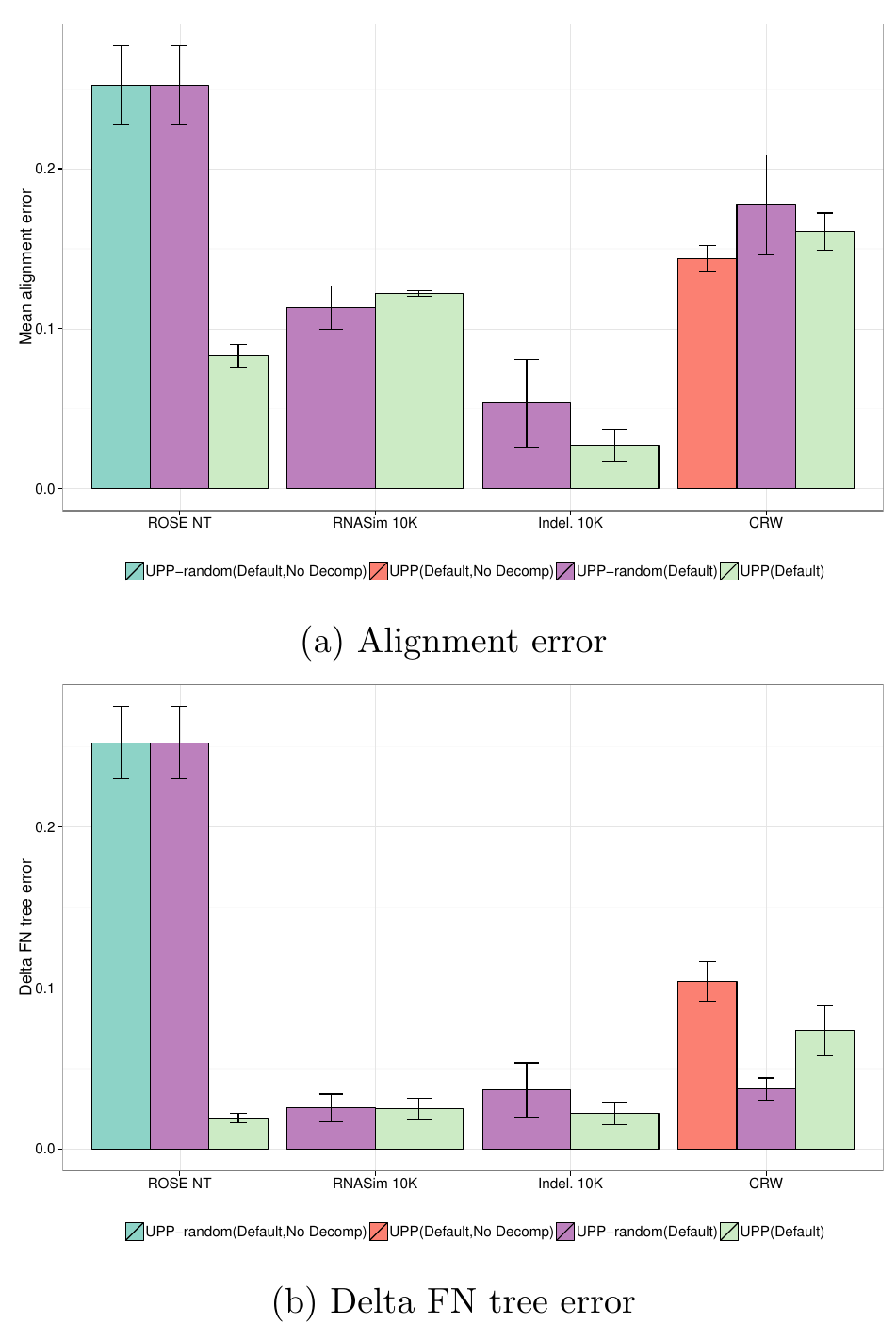}  
\caption[Average alignment SP-error and tree error of UPP variants on the fragmentary datasets.]{\label{frag_upp_hmm}  {\bf Average alignment SP-error and tree error of UPP(Default), UPP(Default, NoDecomp), UPP-random(Default), and UPP-random(Default, NoDecomp) on the fragmentary datasets.}  
UPP-random does not restrict the
backbone to full length sequences, and so allows fragmentary sequences to be in the backbone set.  
UPP-random(Default, NoDecomp) failed to align at least one dataset from each of the RNASim 10K, 
Indelible 10K, and CRW model conditions.  UPP(Default, NoDecomp) failed to align at least one dataset from each of the ROSE NT, RNASim 10K, and Indelible 10K model conditions.  Maximum likelihood trees were estimated using FastTree under GTR.}
\end{figure}

To better understand why UPP is robust to
fragmentation, we explored
UPP variants (called UPP-random) in which we 
did not constrain the backbone to be
only full-length sequences. We also looked
at whether the use of the ensemble of HMMs instead of
a single HMM contributes to robustness to fragmentation.
These comparisons (Fig.~\ref{frag_upp_hmm})
revealed some interesting trends about the impact of
these algorithm design parameters.
First, the only UPP variants that were able to align
all the datasets were the two that used the ensemble
of HMMs; the variants that used a single HMM each
failed to align several datasets because HMMER was not able
to align some of the query sequences to the backbone alignment
(Fig.~\ref{frag_upp_hmm}).
Second, 
 the comparison
between UPP-random(Default) and UPP(Default))
favored UPP(Default), so that
while   there were negligible to small differences
in some cases, UPP(Default) was
dramatically more accurate than UPP-random(Default) on
the ROSE NT datasets for both alignment SP-error
and tree error
(Fig.~\ref{frag_upp_hmm}).
Thus, restricting the backbone to full-length
sequences is a very important contribution to 
robustness to fragmentary sequences. 
However, 
restricting the backbone to full-length sequences 
and using only a single HMM  produced much higher tree error
than using an ensemble of HMMs
(Fig.~\ref{frag_upp_hmm}), 
showing that
using an ensemble of HMMs also provides benefits. 
Thus, the two algorithmic techniques (restricting the backbone to
full-length sequences, and using
an ensemble of HMMs) are both useful to
improving robustness to fragmentary sequences, but they
address different analytical challenges.

\begin{figure}[htpb]
 \centering
 \includegraphics[width=.65\linewidth]{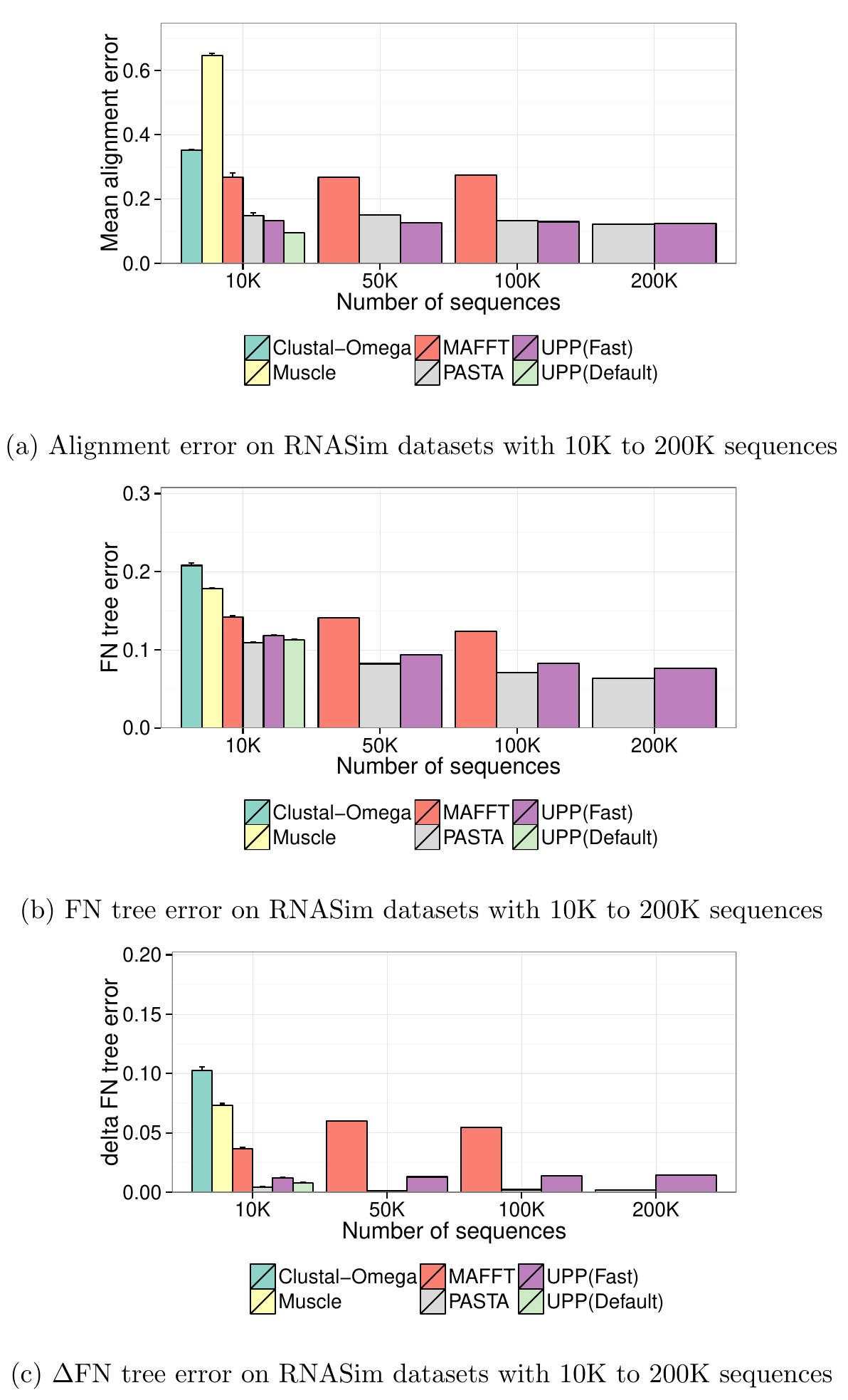} 
 \caption[Alignment SP-error and tree error rates on RNASim datasets with up to
200K sequences.]
 {\label{fig:rnasim_main_large}  {\bf Alignment SP-error and tree error rates on
 RNASim datasets with up to 200K sequences. }
 Results not shown are due to methods failing to return an alignment
 within the 24-hour time period on TACC, using 12 cores.
Maximum likelihood trees were estimated using FastTree under GTR.
 }
 \end{figure}

\paragraph{Impact of taxon sampling. } We evaluated 
the ability for different methods to analyze very
large datasets (up to one million sequences), using subsets of the
million-sequence 
RNASim dataset; this comparison also reveals the impact
of taxon sampling on the alignment methods.
We examined performance for UPP(Fast), the fast version of UPP that differs
from the default setting of UPP only in that it uses
smaller backbones (100 sequences instead of 1000).
Figure \ref{fig:rnasim_main_large} shows
results for 10,000 to 200,000 sequences, and
compares   UPP(Fast), UPP(Default), PASTA, MAFFT, Muscle,
and Clustal-Omega, limiting analyses to 24 hours
on a 12-core 24 Gb machine. 
While all methods shown were able to complete analyses on the 10K dataset,
only UPP(Fast) and PASTA completed analyses on the
100K and 200K datasets.  

As the number of sequences in the RNASim datasets increased, PASTA's alignment SP-error dropped from 15.0\% at 50,000 sequences to 12.2\% at 200,000 sequences.  UPP(Fast) 
had stable alignment SP-error across all the datasets, varying between 12.5 to 13.3\%.  
Both UPP and PASTA trees improved with increased taxon sampling,
with PASTA trees approaching the accuracy of maximum likelihood on the true alignment (0.1\% to 0.2\% $\Delta$FN), and UPP trees close behind (1.2\% to 1.4\% $\Delta$FN,
Fig.~\ref{fig:rnasim_main_large}(c)).

We then compared UPP(Fast) to 
PASTA on the full 1,000,000-sequence RNASim dataset.  
We ran UPP(Fast) and PASTA on a dedicated machine 
with 12 cores and 256 GB of memory so that the 
analyses could exceed the 24 hour time limit in TACC.  
UPP(Fast) completed in 12 days, with alignment and tree errors similar to 
previous results (12.8\% alignment SP-error and 
2.0\% $\Delta$FN). 
PASTA completed in 15 days, and produced a much worse alignment but better tree (18.5\% alignment SP-error and 0.4\% $\Delta$FN).
Because we used a different machine with a different
architecture,  the running
times on the 1,000,000-sequence RNASim dataset cannot be directly compared
to the running times on the other RNASim datasets, which were run on
TACC.

\paragraph{Computational issues. }

Table \ref{table:all_full_time} compares wall
clock running times, using 12 cores,  on those datasets
where all methods were able to complete within the 24 hours 
limitation on Lonestar; thus, we show results on
 all datasets except 
for the RNASim datasets with 50K or more sequences.  
Note that all methods but Muscle had parallel implementations and were able to take advantage of the 12 available cores; the relative performance differences between 
methods could greatly differ on a single core machine, depending on how
well each method is able to take advantage of parallelism.

The differences in average running time on these datasets were sometimes small
(e.g., all methods completed analyses using between 0.4 to 0.6 hours
wall clock time for the ROSE NT datasets with 1000 sequences,
and in less than 0.2 hours wall clock time for the 10 AA datasets
with under 1000 sequences). However,
on the CRW datasets, which could be quite large (nearly 28K sequences),
the differences in average running time were large: UPP(Default)
used 11.6 hours, Muscle used 5.9 hours, PASTA used 3.2 hours,
Clustal-Omega used 2.8 hours, and 
MAFFT used only 0.4 hours. 
Overall, on these datasets, 
MAFFT was generally the fastest (or nearly so), 
and UPP(Default) generally the slowest.

\begin{table}[htbp]
\caption[Wall clock running time across most of the full-length datasets.]{\label{table:all_full_time}
{\bf Wall clock running time across most full-length datasets. }  
We report average wall clock running time  on the full-length datasets for
which most methods could complete; this includes everything other than
the RNASim datasets with 50,000 or more sequences.
UPP is run in default mode.
Results marked with an ``X'' indicate that the method failed to terminate within the time limit (24 hours on a 12 core machine).  All methods but Muscle had parallel implementations and were able to take advantage of the 12 cores.  Muscle failed to align two of the HomFam datasets; 
we report separate average results on the 17 HomFam datasets for all methods and the two HomFam datasets for all but Muscle.
}
\centering
\scalebox{.75}{
\begin{tabular}{|l|r|r|r|r|r|r|r|r|}
  \hline
Method & ROSE  & RNASim  & Indelible  & ROSE  & CRW & 10 AA &HomFam&HomFam\\ 
& NT & 10K & 10K & AA & & &(17) &(2) \\
  \hline
 \multicolumn{9}{|c|}{Average wall clock running time (hr)}\\
  \hline  
  UPP  & 0.6 & 6.7 & 6.7   & 0.2     & 11.6  & $<$0.1  & 1.3    & 0.5\\   
  PASTA         & 0.6 & 3.9 & 1.3   & 0.2     & 3.2   & 0.2     & 1.5    & 1.3\\   
  MAFFT         & 0.4 & 0.1 & 1.4   & $<$0.1  & 0.4   & 0.1     & $<$0.1 & 0.1 \\   
  Muscle        & 0.5 & 0.8 & 1.2   & $<$0.1  & 5.9   & 0.2     & 1.3    & X \\ 
  Clustal & 0.4 & 4.8  & X    & 0.1     & 2.8   & $<$0.1  & 0.3    & 0.3 \\ 
   \hline   
\end{tabular}}
\end{table}

We compared the 
wall clock running time for each stage of the UPP 
algorithm for UPP(Default) and UPP(Fast) on two large
nucleotide datasets: the
RNASim 10K dataset with 10,000 sequences and the CRW 16S.B.ALL dataset
with 27,643 sequences (Table~\ref{runningtime_gutell}).  
Only two steps -- computing the backbone alignment and tree and 
searching for the best HMM -- used more than a few minutes, even
on the largest dataset. Computing the backbone alignment and
tree took under an hour for UPP(Default) and under 8 minutes
for UPP(Fast).
However, searching for the best HMM for
the query sequences took the most time. For UPP(Default),
which had 10 times as many HMMs as UPP(Fast), this
step took
nearly 16 hours  on 16S.B.ALL and 7 hours
on the RNASim 10K dataset, while UPP(Fast) 
used under 1.8 hours on the 16S.B.ALL dataset and
0.8 hours on the RNASim 10K dataset.
Thus, the vast majority of the time on large datasets is spent
searching for the best HMM.
 On very small datasets, the running time 
difference beween UPP(Default) and UPP(Fast) will be small, 
but on very large datasets the running time differences 
will be substantially increased -- close to an 
order of magnitude of difference in running time.

\begin{table}[htpb]
\caption[Running time for UPP stages.]{\label{runningtime_gutell}  
{\bf Running time for UPP(Fast) and UPP(Default) on the RNASim 10K and CRW 16S.B.ALL datasets.}
We show the wall clock running time (hr) for each stage of UPP(Fast) and UPP(Default) on the RNASim 10K (10,000 sequences) and CRW 16S.B.ALL (27,643 sequences) datasets, two of the largest nucleotide datasets.  The UPP alignments were computed on TACC's Lonestar Cluster machine.  The vast majority of the running time was spent searching for the best HMM for the query sequences.}
   \centering
   \scalebox{.75}{
   \begin{tabular}{|l|r|r|r|r|}
   \hline
   &\multicolumn{4}{|c|}{Wall clock running time (hr)}\\
   &\multicolumn{2}{|r|}{RNASim 10K}&  \multicolumn{2}{|r|}{CRW 16S.B.ALL}\\
   \hline
   Stage& UPP(Fast) & UPP(Default)& UPP(Fast) & UPP(Default)\\
   \hline
   Building Backbone & 0.12& 0.42& 0.13& 0.52\\   
   Building HMMs & $<$0.01& 0.02& $<$0.01& 0.02\\
   Searching for best HMM & 0.83& 6.53& 1.81& 15.45\\
   Aligning sequences & 0.02& 0.03& 0.05& 0.15\\
   Merge alignments & 0.01& 0.01& 0.01& 0.02\\
\hline
  {\bf Total time}:  & 0.99 & 7.01 & 2.01 & 16.16 \\
   \hline      
   \end{tabular}}
\end{table}





\begin{figure}[htpb]
\centering
\includegraphics[width=0.8\linewidth]{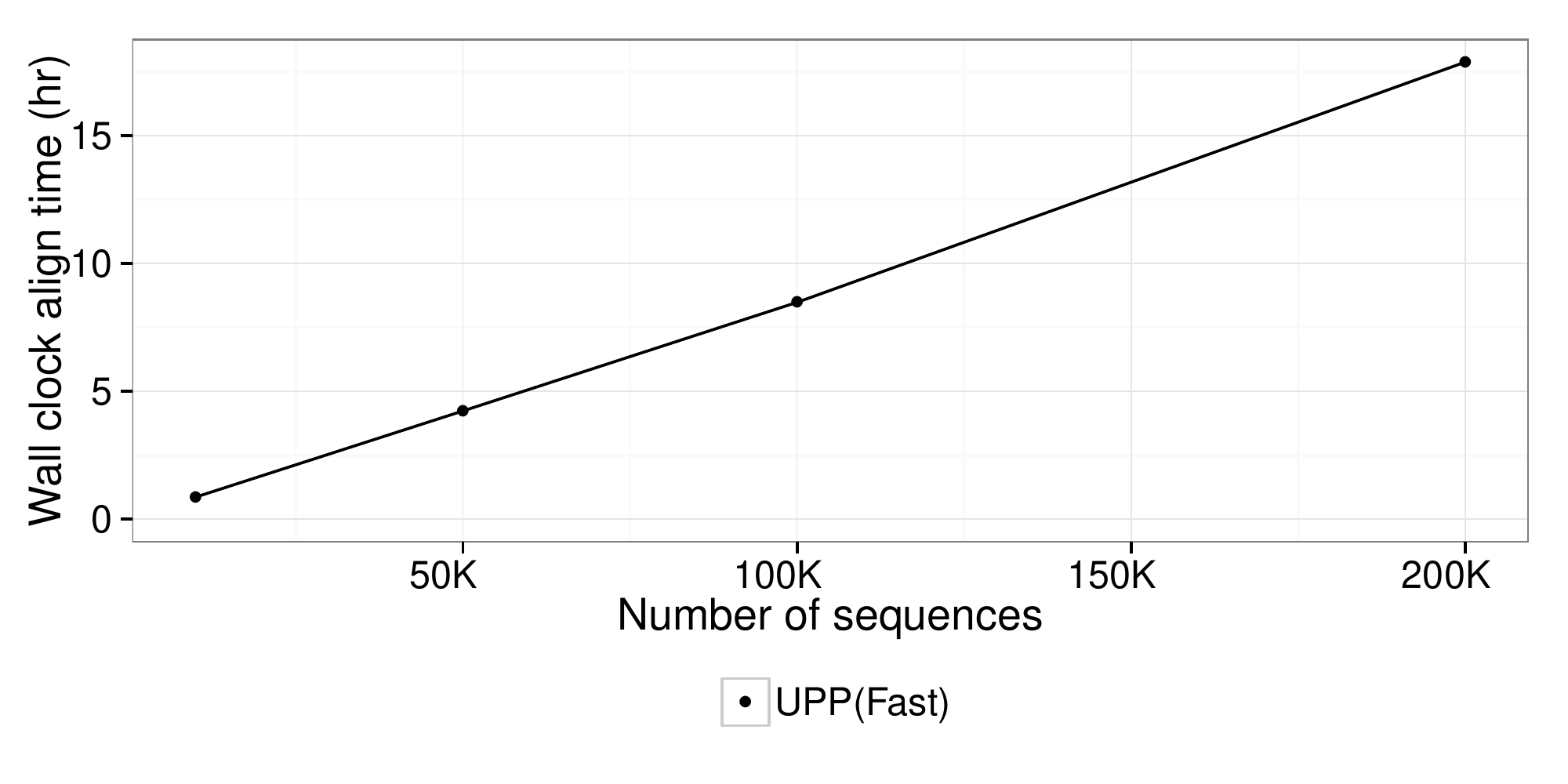}\\
\caption[Running time for UPP(Fast) on the RNASim  datasets.]
{\label{running-time}  {\bf Running time for UPP(Fast) on the RNASim
datasets. }
We show running time to generate an alignment for UPP(Fast) on RNASim datasets
with 10K, 50K, 100K, and 200K sequences.  All analyses were run on TACC with 
24 GB of memory and 12 CPUs.}
\end{figure}

We then explored how UPP's running time (measured using
wall clock time) scaled with the
size of the dataset by exploring subsets of the RNASim dataset
with 10,000 to 200,000 sequences, using 12 cores. 
Running times for UPP(Fast) on the RNASim datasets
showed a close to linear trend, so that
UPP(Fast) completed on 10K sequences in 55 minutes, 
on 50K sequences in 4.2 hours,
on 100K sequences in about 8.5 hours,
and on 200K sequences in about 17.8 hours (Fig.~\ref{running-time}). 

\section*{Conclusions}  

Although the relative  performance
of multiple sequence alignment methods
depended on the  dataset,  in most cases UPP
produced alignments with lower SP-error rates
and higher TC scores
than MAFFT, Muscle, and Clustal-Omega,  and maximum likelihood trees 
computed on UPP alignments were also more accurate than ML trees on these
other alignments. 
However, the comparison between 
UPP and PASTA is more  interesting.
Because UPP uses PASTA to compute its
backbone alignment and tree, 
by design, UPP is identical to PASTA on 
fragment-free datasets containing at most 1000 sequences.
The comparison between UPP and PASTA with respect to
alignment accuracy is interesting:
UPP alignments tend to have lower SP-error  
rates than PASTA alignments but
also lower TC scores, indicating that these two criteria
are not that well correlated.
However, ML trees based on PASTA alignments (for fragment-free datasets) are typically
more accurate than ML trees based on UPP alignments. 
On datasets with fragmentary sequences, UPP has nearly the same
SP-error rates that it achieves on the
full-length sequences,  while PASTA's SP-error rates increase
substantially with fragmentation;
 consequently, 
UPP's 
$\Delta$FN tree error rates do not tend to increase that much with fragmentation
 although they do for PASTA.
Thus, UPP is highly robust to fragmentary data whereas PASTA is not. 
Hence, while PASTA has an advantage over
UPP on datasets without fragments, UPP presents advantages 
relative to PASTA for datasets with
fragments.

To understand UPP's performance, it is useful to consider
the alignment strategy it uses. 
First, it computes a backbone alignment using 
PASTA on a relatively small (at most 1000-sequence) dataset;
this allows it to begin with a highly accurate
alignment. Then, instead of using
a single profile HMM to represent its backbone alignment, 
UPP uses a collection of profile HMMs, each on a subset of the
sequences.
The subsets are obtained from local regions of the backbone tree,
which is an ML tree estimated on the backbone sequences.
Hence,
the sequences in these subsets tend to be closely related.
The induced subset alignments for these smaller localized regions
are thus better suited for HMMs, especially when the full dataset
displays overall substantial heterogeneity.

These observations help explain why
using multiple HMMs, each on a region 
within the backbone tree, 
 provides improved
alignments compared to the use of a single HMM.
However, UPP also 
restricts the backbone to the full length sequences, and
this algorithmic step is critical to improving
robustness to fragmentary
sequences. 
Hence, these aspects of UPP's algorithmic design -- restricting
the backbone to full length sequences and using an ensemble of 
HMMs instead of a single HMM -- increase sensitivity to
remote homology (especially for fragmentary sequences)
and reduces alignment 
SP-error
and tree error, but
each targets a different
aspect of the algorithmic performance.

UPP exhibits great scalability with respect to
running time (which scales in a nearly linear manner),
parallelism, 
 and alignment accuracy.
For example, our study showed the alignment 
SP-error on the backbone alignment is quite 
close to the alignment SP-error on the alignment
returned by UPP. 
Thus, UPP enables
large datasets to be aligned nearly as accurately as smaller datasets.  

Overall,  UPP is a multiple sequence alignment
method that can provide very high accuracy on
sequence datasets that have been considered too difficult
to align, including
datasets that evolved with high
rates of evolution, that
contain fragmentary sequences, or that contain
many thousands of sequences - even up to one million
sequences. 
UPP performs well on
both phylogenetic and structural benchmarks
(see
\cite{ReeckMuddle}  for further discussion of these 
related but different tasks).
Finally, 
UPP is parallelized (for shared memory) and has a checkpointing feature, but does not require supercomputers
to achieve excellent accuracy on ultra-large datasets in reasonable timeframes.

\section*{Methods}

\subsection*{Performance Study}\label{performance_study}

\paragraph{Data and software availability}
The datasets used in this study are available at~\cite{upp_submission_site}.
The github site for UPP
\cite{sepp_github}
provides  open source software and  
instructions on how to
download, install, and run UPP.

\paragraph{Datasets.} 
All datasets used in our study were used
in 
previously published studies, and
are available online through the respective publications.
Because UPP is designed for ultra-large scale multiple
sequence alignment, we focus
the analysis on benchmark datasets with many sequences.
We used the following collections
of  simulated datasets:
\begin{itemize}
\item
{\bf ROSE NT:}  a
collection of 1000-sequence nucleotide datasets 
from \cite{Liu2009} that were
generated using ROSE \cite{Stoye1998}; see
\cite{Liu2009} for full details.
\item
 {\bf Indelible 10K:}  a collection of
10,000-sequence 
nucleotide datasets from \cite{pasta} 
that were generated by Indelible \cite{Fletcher01082009}; see \cite{pasta} for full details.
\item
{\bf 
RNASim:}  a collection of
datasets ranging from 
 10,000 sequences to 1,000,000 sequences
\cite{pasta-jcb}. 
\item
{\bf ROSE AA:} a collection of 
5000-sequence simulated AA datasets 
from \cite{Price2010} that were
generated using ROSE.
\end{itemize}

\noindent
We also used biological datasets 
with reference alignments that were used
in prior studies \cite{Liu2012,pasta-jcb,Sievers2011} to evaluate
alignment methods on large datasets. 
We focus on datasets with 10,000 or more sequences, but
also used ten large amino-acid datasets (8 from the BAliBASE
\cite{Thompson2011} collection, and two others) with at least 320 
sequences. 
\begin{itemize}
\item
{\bf  CRW:}  the three largest 
datasets from the
Comparative Ribosomal Website (CRW)  \cite{Cannone2002}, each
a set of 16S sequences. We include
the 16S.3 dataset (6,323 sequences spanning three phylogenetic domains), the 
16S.T dataset (7,350 sequences spanning three phylogenetic domains), 
and the 
16S.B.ALL dataset (27,643 sequences spanning the bacteria domain).  
The CRW datasets have highly reliable, curated alignments inferred 
from secondary and tertiary structures and were previously studied in~\cite{liu-ploscurrents,Liu2012}.
The reference trees on these datasets 
used in these studies 
were derived from maximum
likelihood trees estimated using RAxML, with all
branches with bootstrap support below 75\% 
collapsed.
\item
{\bf 10 AA:}  ten
amino acid
datasets  with curated multiple sequence alignments (the eight 
largest
BAliBASE datasets \cite{Thompson2011}
and IGADBL\_100 and coli\_epi\_100 from \cite{Gloor2005}); these
range in 
size from
 320 to 807 sequences, and were used
in \cite{pasta-jcb} to evaluate multiple sequence
alignment methods.
The reference trees on these datasets used
in these studies were based on
RAxML with all branches with bootstrap support below 75\%  collapsed.
\item
{\bf HomFam:}  a collection of 
19 of the largest HomFam datasets, which are amino acid
sequence datasets ranging in size from 
10,099 to 93,681 sequences with Homstrad \cite{homstrad} reference
alignments on small subsets (5-20 sequences, median 7).
These 19 datasets 
were used in \cite{Sievers2011,pasta-jcb} to evaluate  multiple
sequence alignment methods on large amino acid datasets.
The study in \cite{Sievers2011} also explored performance on 
smaller HomFam datasets, but these
are not as relevant to this study. As noted in \cite{pasta-jcb}, the HomFam rhv 
dataset studied in \cite{Sievers2011} had a warning
on the PFAM website that the alignment was ``very weak"; for that
reason, the rhv dataset
was omitted from the study reported in \cite{pasta-jcb} and from this one.
\end{itemize}

For some of the nucleotide datasets, 
we generated three fragmented versions, 
making 12.5\%, 25\%, and 50\% of the
sequences fragmentary.  The lengths of the fragments were drawn from 
a normal distribution with a mean length of 500 bps and a standard deviation of 60 bps (mean length is one-third of the average length of the CRW datasets and one-half the length of the Indelible and ROSE NT datasets).  
We generated fragmentary datasets by selecting a random subset 
of sequences and a random substring (of the given length) for each selected sequence.

\paragraph{Alignment and Tree Estimation Software. }

\subsection*{Basic alignment methods}\label{commands}
Each dataset was aligned (when possible) 
using Clustal-Omega \cite{Sievers2011} version 1.2.0, MAFFT \cite{Katoh2007} version 6.956b, MUSCLE \cite{Edgar2004a} version 3.8.31, and PASTA version 1.5.1 \cite{pasta,pasta-jcb}.  
MUSCLE was run with the ``-maxiters 2'' option on datasets of 3000 sequences or greater.  
Due to a bug in earlier versions of MAFFT 6.956b, MAFFT-default was run using MAFFT version 7.143.  
We ran three different versions of MAFFT.  MAFFT-L-INSI was run on datasets with 1,000 for fewer sequences.  
For most datasets with more than 1,000 sequences, we ran MAFFT-default (``-{}-auto''); the exceptions were
the RNASim 100K dataset, three replicates from the Indelible 10K 10000M3 dataset, and the CRW 16S.B.ALL dataset,
where MAFFT-default  failed to run and so we used
MAFFT-PartTree.
All MAFFT variants included the ``-{}-ep 0.123'' parameter.

Because the algorithmic design parameters for running PASTA on amino acid
datasets had not been studied, we examined different
options for running 
PASTA on amino acid datasets 
and used those settings in our studies of amino acid datasets
(see SOM Section \ref{pasta_protein}).
PASTA was run for three iterations
or a maximum of 24 hours, whichever came first.  
If PASTA did not terminate
at the end of 24 hours, the alignment
from the last successfully completed iteration were used.
PASTA was run using a MAFFT-PartTree starting tree for all but the RNASim datasets.  
For the RNASim datasets, we used the ML tree estimated on the UPP(Fast, NoDecomp) alignment 
as the starting tree (MAFFT-PartTree was unable to run on the largest RNASim datasets).  
The remaining settings for PASTA were set using the ``-{}-auto'' flag.  

Commands for each method are given below:
\begin{itemize}
\item \textbf{Clustal-Omega}:~\emph{clustalo -{}-threads=12 -i$<$input\_sequence$>$ -o $<$output\_alignment$>$}
\item \textbf{MAFFT-L-INS-i}:~\emph{mafft  -{}-ep 0.123 -{}-thread 12 -{}-localpair -{}-maxiterate 1000 -{}-quiet -{}-anysymbol $<$input\_sequence$>$ $>$ $<$output\_alignment$>$}
\item \textbf{MAFFT-default}:~\emph{mafft  -{}-thread 12 -{}-ep 0.123 -{}-auto -{}-quiet -{}-anysymbol $<$input\_sequence$>$ $>$ $<$output\_alignment$>$}
\item \textbf{MAFFT-PartTree}:~\emph{mafft -{}-thread 12 -{}-ep 0.123 -{}-parttree -{}-retree 2 -{}-partsize 1000 -{}-quiet $<$input\_sequence$>$ $>$ $<$output\_alignment$>$}
\item \textbf{MAFFT-profile}:~\emph{mafft [-{}-localpair -{}-maxiterate 1000] [-{}-addfragment $\vert$ -{}-add] $<$query\_file$>$ $<$backbone\_alignment$>$ $>$ $<$output\_alignment$>$}
\item \textbf{MUSCLE}:~\emph{muscle [-maxiters 2] -in $<$input\_sequence$>$ -out $<$output\_alignment$>$}
\item \textbf{PASTA}:~\emph{python run\_pasta.py  -{}-num-cpus=12 -o $<$output\_directory$>$ -i  $<$input\_sequences$>$ -t $<$starting\_tree$>$ -{}-auto -{}-datatype=$<$molecule\_type$>$}
\item \textbf{UPP}:~\emph{python exhaustive\_upp.py -a $<$backbone\_alignment$>$ -t $<$backbone\_tree$>$ -s $<$query\_sequences$>$ -d $<$output\_directory$>$ -o $<$output\_name$>$ -x 12 -A 10 -m $<$molecule\_type$>$ -c $<$default\_config\_file$>$}
\item \textbf{UPP-disjoint}:~\emph{python exhaustive\_upp.py -S normal -a $<$backbone\_alignment$>$ -t $<$backbone\_tree$>$ -s $<$query\_sequences$>$ -d $<$output\_directory$>$ -o $<$output\_name$>$ -x 12 -A 10 -m $<$molecule\_type$>$ -c $<$default\_config\_file$>$}
\end{itemize}

\subsection*{HMMER Commands}\label{hmm_commands}
HMMER 3.0~\cite{HMMER} was used internally 
within UPP for building the ensemble of HMMs (hmmbuild), 
for searching for the best HMM for a query sequence (hmmsearch), and 
for inserting the query sequence into the alignment (hmmalign):

\begin{itemize}
\item \textbf{hmmbuild}:\\\emph{hmmbuild -{}-symfrac 0.0 -{}-informat afa -{}-$<$molecule\_type$>$ $<$output\_profile$>$ $<$backbone\_alignment$>$}
\item \textbf{hmmsearch}:\\\emph{hmmsearch -{}-noali -o $<$output\_file$>$ -{}-cpu 1 -E 99999999 -{}-max $<$input\_profile$>$ $<$query\_file$>$}
\item \textbf{hmmalign}:\\\emph{hmmalign -{}-allcol -{}-dna $<$output\_profile$>$ $<$query\_file$>$ $<$output\_alignment$>$}
\end{itemize}

\subsection*{Maximum Likelihood Tree Estimation}\label{ml_estimation}

%
To compute maximum likelihood trees on
large datasets (with 1000 or more sequences)   we used 
FastTree \cite{Price2010} version 2.1.5 SSE3, and  we used RAxML 
\cite{Stamatakis2014} version 8.0.6 for smaller datasets. 
We used the General Time Reversible (GTR) model for
all the  nucleotide datasets (simulated
and biological) and JTT for the simulated amino acid datasets
(ROSE AA). For the 10 AA datasets (all biological), we
used 
ProtEST \cite{Cuff2000} to select the 
model for each dataset,
and then used that model within RAxML to perform
the analysis. 
The version number and commands used to run each method are given below.

\begin{itemize}
\item \textbf{FastTree AA}:\\\emph{FastTreeMP -nosupport $<$input\_fasta$>$ $>$ $<$output\_tree$>$}
\item \textbf{FastTree NT}:\\\emph{FastTreeMP -nosupport -nt -gtr $<$input\_fasta$>$ $>$ $<$output\_tree$>$}
\item \textbf{RAxML AA}:\\\emph{raxmlHPC -T 12 -m PROT $<$model\_name$>$GAMMA -j -n $<$output\_name$>$ $<$starting\_tree$>$ -s $<$input\_fasta$>$ $>$ -w $<$output\_directory$>$ -p 1}
\end{itemize}

\paragraph{Performance Metrics.}
We compare estimated alignments and their ML trees to reference 
alignments and trees.
We use FastSP \cite{fastsp} to 
compute SP-error (the average of SPFN and SPFP error) and TC scores. 
The SPFN rate is the
 sum-of-pairs false 
negative  rate (which is the percentage of the
homologous pairs in the reference alignment that
are not in the estimated alignment) and the
SPFP is the
sum-of-pairs false positive  rate
(which is the
percentage of homologous 
pairs in the estimated alignment that are not present in the reference alignment).  

We report tree error using the false negative (FN) rate (also
known as the missing branch rate), which is 
the percentage of internal edges in the reference tree that 
are missing in the estimated tree.
We also report $\Delta{FN}$, the difference between 
the FN rate of the estimated tree and the FN 
rate of the tree estimated on the true alignment, to 
evaluate the impact of alignment estimation on
phylogenetic analysis.  
Most typically, $\Delta{FN}>0$, indicating that the estimated tree
has higher error than the ML tree on the true alignment, but
it is possible for $\Delta{FN}<0$, which happens when  
the estimated ML tree is more accurate than ML on the
true alignment.

\section*{Acknowledgements}
The authors thank the Texas 
Advanced Computing Center (TACC) at The University of Texas at Austin for
providing HPC resources that contributed to the research results reported within this paper. TW was
supported by the U.S.~National Science Foundation  (NSF) grants
0733029 and 1461364; SM was supported by an international
predoctroal fellowship from the Howard Hughes Medical
Institute (HHMI); and NN was supported by the University of 
Alberta through a grant to TW and by NSF grant 1461364.
The authors thank Erich Jarvis,
Tom Gilbert, Jim Leebens-Mack,
Ruth Davidson, Michael Nute, and the anonymous reviewers for 
their helpful critiques of early versions of the manuscript.

\bibliographystyle{model1a-num-names}
\bibliography{upp-new-2}

\end{document}